# Stabilization Based Networked Predictive Controller Design for Switched Plants

Avijit Routh[1,2], Sourav Das[1],
1. Department of Power Engineering, Jadavpur University, Salt-Lake Campus, LB-8, Sector 3, Kolkata-700098, India.
2. Department of Electronics and Instrumentation Engineering, National Institute of Science and Technology, Palur Hills, Berhampur-761008, Orissa, India.

Saptarshi Das[1,3], Indranil Pan[1]
3. Communications, Signal Processing and Control Group, School of Electronics and Computer Science, University of Southampton, SO17 1BJ, Southampton, UK.
Email: saptarshi@pe.jusl.ac.in, s.das@soton.ac.uk

*Abstract*—**Stabilizing state feedback controller has been designed in this paper for a switched DC motor plant, controlled over communication network. The switched system formulation for the networked control system (NCS) with additional switching in a plant parameter along with the switching due to random packet losses, have been formulated as few set of non-strict Linear Matrix Inequalities (LMIs). In order to solve non-strict LMIs using standard LMI solver and to design the stabilizing state feedback controller, the Cone Complementary Linearization (CCL) technique has been adopted. Simulation studies have been carried out for a DC motor plant, operating at two different sampling times with random switching in the moment of inertia, representing sudden jerks.**

## I. Introduction

Networked control systems are distributed control systems where sensors, actuators etc. are interconnected through a real time network and exchange information among themselves through that common shared network. NCSs are widely used in spacecraft control, robotics, process control, factory automation etc. [1]-[2]. NCS reduces wiring and makes installation simpler. It is easy to maintain and also makes analysis of the system properly. But in NCSs, random time delay and packet drop-outs are the major causes which detonate the system performance and leads to system's instability [3]-[4]. In safety critical applications, often the time driven controllers are preferred instead of the event driven controller due to the ease of hardware verification [5]. Thus for event-driven controllers random time-delays, due to the introduction of the communication network in the control loop reduces to packet drop only which is assumed in this paper, for mathematical formulation of stability and controlling a DC motor plant whose load is varied arbitrarily within few discrete set of values. In NCSs several methodologies, such as stochastic control, predictive control, robust control and memory less state feedback control (MSFC) are used to enhance the performance of the system and also compensate the random packet losses and time delay [6]. Here in this paper we use a predictive networked control scheme proposed by Pan *et al.* [7] to control a plant whose one parameter is arbitrary switching and can take few set of discrete pre-specified values. Here, we consider a DC motor plant whose load is changing abruptly within few discrete values and therefore one of the motor parameters i.e. moment of inertia ($J$) also changes accordingly [8]-[9].

Generally, the NCS can be analyzed as a set of Linear Time Invariant (LTI) systems with arbitrary switching between them caused by random data loss and variable delay. This mechanism avoids the problem of formulating it as an asynchronous dynamical system (ADSs) with variable matrix size for stability analysis. But in ADS, there is no standard solver for the Bi-linear Matrix Inequality (BMI) constrains in polynomial time. But for stability analysis of switched LTI systems, the equations can be formulated by LMI constrains which can be solved using standard LMI solvers and the stability can be guaranteed in spite of arbitrary switching (due to packet drops and random delays) in the sense of Lyapunov. Also, other control methods can be augmented to meet better control performance within the LMI framework by easily augmenting the individual LMIs.

In stabilization problems in networked control systems with random packet losses and variable delays, often the Lyapunov analysis gives a set of non-strict LMIs as in [10]-[12] which can not be solved directly with standard LMI solvers. The non-strict LMIs arising from stability condition for switched systems can be solved using the Cone Complementary Linearization based sequential optimization technique subjected to LMI constraints as in Ghaoui *et al.* [10]. Also, in the present paper we used the predictive state-feedback controller for stabilization of the NCS instead of simple state-feedback controller as proposed in Pan *et al.* [7]. This typical controller is capable of producing greater robust stability for the nominal system (in the absence of random packet losses) over network with packet drops. It is based on a priori selection of the future control signals and the associated state feedback gains if the control packets in the forward path get dropped. This particular scheme of predictive state feedback control (Fig. 1) is capable of providing greater stability region in the controller parameter space compared to the classical state feedback control scheme [7]. Here, the future control signals are predicted a priori and fused together with the current one and finally sent to the buffer, to handle any arbitrary drop in future two samples. This paper extends the concept of using networked predictive controller to handle arbitrary parameter switching in the plant itself along with the switching due to the random packet losses in the network.

In this study, our aim is to deal with the stabilization problem of a plant (DC motor) over NCS with random packet losses and having random switching in one additional plant parameter (moment of inertia). Here, we design a state feedback controller and stability analysis approaches in discrete time domain are carried out, while considering random packet drop-outs. The stability conditions for NCSs are formulated in terms of few set of non-strict LMIs which are then solved using the above discussed CCL technique of sequential optimization with LMI constraints, using standard LMI solvers [13] to produce stabilizing predictive controller gains for the switched plant over communication network.

The remainder of the paper is organized as follows. The formulation of the stability condition for NCSs with packet losses is given in section 2. The stabilization based state feedback controller design problem is formulated in section 3. Simulation results for a switched load DC motor plant are demonstrated in section 4. The paper ends with conclusion as section 5, followed by the references.

## II. PROBLEM FORMULATION

As shown in Fig 1, the NCS can be divided into four modules and each part is described in the following sections.

*1) The plant and the sensor*

*2) The data network*

*3) The network controller*

*4) The actuator with buffer*

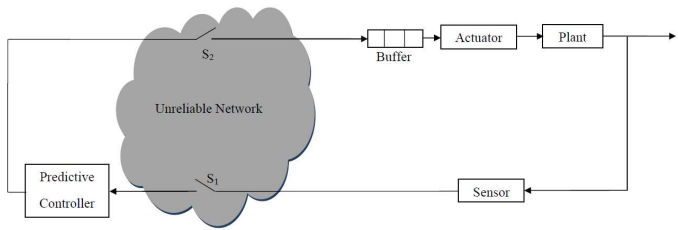

Figure 1. Schematic diagram of networked predictive control system with a plant whose parameter randomly varies within a specified set of values.

### A. Controlled Plant and Sensor

Let us Consider, a plant whose single parameter is changing randomly and can take any of the $r$ different discrete values. The dynamics of the controlled plants are given by the following linear discrete model:

$$x_l(k+1) = F_l x_l(k) + G_l u_l(k), l = 1, \ldots, r \quad (1)$$

where $x_l(k) \in \mathbb{R}^n$ are the plant's state and $u_l(k) \in \mathbb{R}^m$ are the control inputs. $l$ denotes index number of state space of the plant when the parameter is changed to $l^{th}$ value. $F_l$ and $G_l$ are the discrete time system matrices with appropriate dimensions. Note that (1) can be considered as the discretized form of a continuous-time system given by

$$\dot{x}_l(t) = A_l x_l(t) + B_l u_l(t) \quad (2)$$

with sampling period $h$ and therefore

$$F_l = e^{A_l h}, G_l = \int_0^h e^{A_l \tau} d\tau B_l \quad (3)$$

The sensors are assumed to be time driven. At each sampling period, the sampled plant states are encapsulated into a packet and sent to the controller via the network along with the time-stamp (i.e., the time at which the plant's output state is sampled).

### B. Data Network

In practice, the data packets in NCSs generally suffer random time delay and packet losses during network transmission. Here we only consider random packet drops for simplification of the problem and also due to the verification issues of event driven controllers in safety critical applications. The switches $S_1$ and $S_2$ are used to represent the packet losses in the backward and forward networks respectively as shown in Fig. 1. Opening of switch $S_1$ denotes that data packets are lost during the transmission from sensor to controller side. Whereas closing of switch $S_2$ indicates successful transmission of data packet from controller to actuator side.

### C. Networked Controller

The state feedback controller is assumed to be time driven. Generally the convention is to use an event driven controller which calculates the control signal as soon as it is receives the packet and sends it to the actuator via the shared network. This eliminates the additional time taken for the next time trigger. But in safety critical applications the even driven controller is not used because after long delay when many control signal come together at a same instant, the network load increases to a high value which is undesirable. The predictive controller proposed in this paper is a full state-feedback one with the following form:

$$u_l = K_q x_l, \quad q \in \{1, 2, \cdots, N_{drop}\} \quad (4)$$

where, $K_q \in \mathbb{R}^{m \times n}$ is the state feedback gain which has to be designed.

The predictive state feedback controller calculates the values of the present control signal and the next predicted control signals for the plant, keeping in mind that there might be arbitrary packet losses in future two sampling instants which should be compensated by the predicted control signal. The controller modifies the control signals for the plant using the state feedback gains which are the design variables in the present case. Then the data (with present control signal and future control signals for two consecutive samples) are encapsulated in a packet and sent to the buffer.

### D. Actuator

The actuator is considered to be time driven. Here it is assumed that the actuator and sensor have the same sampling time. Though the actuator and the sensor are in same side of the switches, it is easy to synchronize these two elements via hardware synchronization. In this paper, we neglect the variable delays and only consider the packet drop. There is a skew between the time line of the controller, sensor and the actuator. If $\tau_1$ time is taken by sensor to send the data to the controller side and $\tau_2$ time is taken by the controller to send the data packet to the actuator side, then the following assumption to be hold:

$h > \tau_1 + \tau_2$, where $h$ is the sampling time of the system. Thus the timelines have an offset which is assumed to be maintained with accuracy so that controller as well as the actuator can pick up the current signal properly. In practical situation so many control loops use the same shared network which introduces random packet delay. However a packet which belongs to a particular time step is considered to be dropped if it does not reach the actuator in the same time step. Here, any out of order packet is treated as dropped packet.

### E. Modeling of NCSs

According to our assumption, in the NCS, a sensor packet belonging to a particular time step is said to be effective packet, if it is transmitted from the sensor to the controller and corresponding controller input is transmitted from the controller to the actuator successfully in the same time stamp. Let, $S \triangleq \{i_1, i_2, i_3, \cdots\}$, a subsequence of $\{0,1,2,3,\cdots\}$, denote the sequence of time indexes of effective packets. Therefore the packets between the two effective sensor packets can be considered as dropped packets. Based on the above notion, the packet loss in NCSs can be defined as

$$\{\eta(i_m) \triangleq i_{m+1} - i_m, i_m \in S\} \quad (5)$$

which denotes that, from $i_m$ to $i_{m+1}$, the number of dropped packets is $\eta(i_m) - 1$. Also, let $N_{drop} := \max_{i_m \in S} \{\eta(i_m)\}$ be the maximum upper bound of consecutive packet losses. Then we can easily concluded that $\eta(i_m)$ takes values in a finite set $\Omega \triangleq \{1, 2, \ldots, N_{drop}\}$. The packet loss process is assumed to take random values in $\Omega$.

The control signal can be represented as $u_{pq} = K_q x(i_p)$, where $K_q = [K_{q1} \ K_{q2} \cdots K_{qn}]$, $n$ represents number of state variables, $p \in \{1,2,3,...\}$ denotes the time stamp of effective packets and $q \in \{1, 2, 3, \cdots, N_{drop}\}$ represents the index numbers of the next $N_{drop}$ number of predicted control signals from the present time stamp.

### F. Buffer logic

The buffer is a memory device which registers $N_{drop}$ number of data at every sampling instant in an array. The buffer logic is such that, if there is no drop, it will send the control input $u_{p1}$ at $p^{th}$ instant. It will send $u_{p2}$ at $(p+1)^{th}$ instant if $(p+1)^{th}$ packet is dropped, $u_{p3}$ at $(p+2)^{th}$ instant if $(p+2)^{th}$ is dropped and so on. Detail of buffer logic is described in [7].

### III. STABILIZING USING NETWORKED PREDICTIVE CONTROLLER WITH PACKET LOSSES

#### A. Stability Analysis Of Closed Loop NCSs

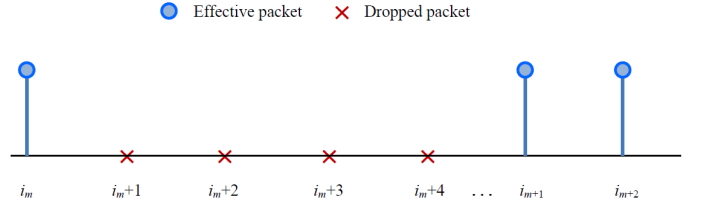

Figure 2. Illustration of packet drops in the predictive NCS.

According to Fig. 2, the NCS dynamics can be described as

$$x_l(i_m + 1) = F_l x_l(i_m) + G_l K_1 x_l(i_m) \quad (6)$$

$$x_l(i_m + 2) = F_l x_l(i_m + 1) + G_l K_2 x_l(i_m) \\ = F_l^2 x_l(i_m) + F_l G_l K_1 x_l(i_m) + G_l K_2 x_l(i_m) \quad (7)$$

In this technique, for time instant $i_{m+1}h$ the NCS dynamics can be written as

$$x_l(i_{m+1}) = F_l^{\eta(i_m)} x_l(i_m) + \sum_{t=0}^{\eta(i_m)-1} F_l^{\ j} G_l K_{\eta(i_m)-t} x_l(i_m) \quad (8)$$

Now, we introduce the augmented state

$$\Gamma_l(i_m) = \left[ x_{l,i_m}^T, x_{l,i_{m-1}}^T, x_{l,i_{m-2}}^T, \ldots, x_{l,i_{m-N_{drop}+1}}^T \right]^T$$

$$\Gamma_l(i_{m+1}) = \Phi_{l,\sigma(i_m)} \Gamma_l(i_m) \quad (9)$$

where, $\sigma(i_m)$ is a piecewise constant function called a switching signal, which takes value in a finite set $\Omega \triangleq \{1, 2, \cdots, N_{drop}\}$. Therefore,

$$\Phi_{l,\sigma(i_m)} = \begin{bmatrix} F_l^{\eta(i_m)} + \sum_{t=0}^{\eta(i_m)-1} F_l^{\ n} G_l K_{\eta(i_m)-t} & 0 & 0 & \cdots & 0 \\ I & 0 & \cdots & \cdots & 0 \\ 0 & I & 0 & \vdots & \vdots \\ \vdots & & \ddots & 0 & \vdots \\ 0 & & \cdots & 0 & I & 0 \end{bmatrix} \quad (10)$$

#### B. Theorem 1:

The system (9) will be asymptotically stable for the arbitrary packet losses if there exists symmetric positive definite matrix $P \in \mathbb{R}^{nN_{drop} \times nN_{drop}}$ satisfying the linear matrix inequalities (LMIs)

$$\Phi_{l,i}^T P \Phi_{l,i} - P < 0 \quad (11)$$

where, $\Phi_{l,i}$ is of the form (10) with specified controller gains $K_q \forall q \in \Omega$.

*Proof:* For the switched system (9) let us define a state depended quadratic Lyapunov function of the form (12),

$$V(i_m) = \Gamma_l^T(i_m) P \Gamma_l(i_m) \quad (12)$$

where, $P$ is symmetric positive definite matrix.

Let the value of $\sigma$ at the $i_m h^{th}$ time instant be $i$. The difference of the Lyapunov function between the two instants of time is given by:

$$\begin{aligned}\Delta V(i_m) &= V(i_{m+1}) - V(i_m) \\ &= \Gamma_l^T(i_{m+1}) P \Gamma_l(i_{m+1}) - \Gamma_l^T(i_m) P \Gamma_l(i_m) \\ &= \Gamma_l^T(i_m)(\Phi_i^T P \Phi_i - P)\Gamma_l(i_m)\end{aligned}$$

For any $\Gamma_l(i_m) \neq 0$, $\Delta V(i_m) < 0$ if (11) holds.

Thus, $\lim_{i_m \to \infty} V(i_m) = 0$. Hence the system (9) is asymptotically stable. As the parameter varies in a discrete manner, the system can be viewed as if its state matrices are also switched randomly within a range and can take few pre-specified value.

### C. Theorem 2:

It can be seen from equation (11) that it is non-linear on the variables of $K_q$ and $P$. So a stabilizing controller cannot be easily obtained directly from the conditions (11). But equation (10) can be made solvable if there exist symmetric positive definite matrices $P \in \mathbb{R}^{nN_{drop} \times nN_{drop}}$ and $Q \in \mathbb{R}^{nN_{drop} \times nN_{drop}}$ and gain matrices $K_q \in \mathbb{R}^{1 \times n}$ such that

$$\begin{bmatrix} -P^T & \Phi_{l,i}^T \\ \Phi_{l,i} & -Q \end{bmatrix} < 0 \quad (13)$$

$$PQ = I \quad (14)$$

where, $\Phi_{l,i}$ is of the form (9) with specified controller gains $K_q$.

It can be noticed that conditions in theorem 2 are not the strict LMI form due to eq. (14). By using cone complementarily linearization (CCL) approach as in [10]-[12], we can solve this non-convex feasibility problem into a sequential optimization problem subject to LMI constrains. The basic idea of CCL algorithm is that, if the LMI

$$\begin{bmatrix} P & I \\ I & Q \end{bmatrix} \geq 0 \quad (15)$$

is feasible with the matrix variables $P \in \mathbb{R}^{n \times n} > 0$ and $Q \in \mathbb{R}^{n \times n} > 0$, then $trace(PQ) = n$ if and only if $PQ = n$. Hence, it is possible to solve equation (13) by applying the CCL algorithm.

### D. Single Controller Design for a Plant Whose Load is Varying in a Discrter Manner:

The aim of this paper is to design a single controller which can stabilize arbitrary load changes in the plants. For that, a common set of controller gain for those plants is to be found. Hence for $r$ number of different loads, common $P, Q$ and $K_q$ matrices have to be found.

*Step 1:* Find a feasible set $(P(0), Q(0), K_q(0))$ satisfying (10), (13) and (15) for $l = 1, \cdots, r$. Set $k = 0$.

Step2: Solve the following LMI problem:

OP:

$$\min \quad trace(PQ(k) + P(k)Q)$$

Subject to inequalities (10), (13) and (15) for $l = 1, \cdots, r$.

*Step3:* Substitute the obtained matrix variables $(P, Q, K_q)$ in (10) and (11). If (10) and (11) are satisfied for $l = 1, \cdots, r$, with

$$|trace(PQ(k) + P(k)Q)| < \partial$$

where, $\partial > 0$ is a sufficient small scale, then output $(P, Q, K_q)$. Exit.

*Step4:* If $k > N$, where $N$ is the maximum number of iterations allowed, exit.

*Step5:* Set $k = k+1$, $(P(k), Q(k), K_q(k)) = (P, Q, K_q)$ and go to step 2.

## IV. SIMULATION AND RESULTS

For simulation study a DC motor plant has been considered whose parameters are given in Table 1 [8]-[9].

TABLE I. PARAMETERS OF THE DC MOTOR PLANT

| | | |
|---|---|---|
| $J$ | Inertia | 0.03, 0.02, 0.01 Kg.m$^2$ |
| $L$ | Inductance | 0.5 H |
| $R$ | Resistance | 2 Ω |
| $K_m$ | Torque Constant | 0.15 N.m/A |
| $\overline{B}$ | Damping Coefficient | 0.2 N.m.sec/rad |
| $K_b$ | Back-EMF Constant | 0.15 V.s/rad |

In this paper the DC motor system can be described as

$$\begin{aligned}\dot{i}_a &= -\frac{R}{L} i_a - \frac{K_b}{L} \omega + \frac{1}{L} u \\ \dot{\omega} &= \frac{K_t}{J} i_a - \frac{\overline{B}}{J} \omega\end{aligned} \quad (16)$$

where $i_a$ is the armature winding current; $\omega$ is the rotor angular speed; $R$ is the armature winding resistance; $L$ is the armature winding; $K_b$ is the back-electromotive-force (EMF) constant; $u$ is the armature winding input voltage; $K_t$ is the torque constant; $J$ is the system moment of inertia; $\bar{B}$ is the system damping coefficient. If the load of the plant (DC motor) is changed arbitrarily within few discrete set of values, then the motor parameter $J$ will be changed significantly. Assuming the state variables as $x \triangleq [i_a, \omega]^T$, the DC motor system can be expressed by the following state-space (17) with the system matrices given by (18)

$$\dot{x}(t) = Ax(t) + Bu(t)$$
$$y(t) = Cx(t) + Du(t)$$
(17)

$$A = \begin{bmatrix} -\dfrac{R}{L} & -\dfrac{K_b}{L} \\ \dfrac{K_m}{J} & -\dfrac{\bar{B}}{J} \end{bmatrix}; B = \begin{bmatrix} \dfrac{1}{L} \\ 0 \end{bmatrix}; C = \begin{bmatrix} 0 & 1 \end{bmatrix}; D = [0]$$
(18)

where, $y(t)$ is the rotor angular speed. Hence, state space models of the DC motor can be obtained by varying the inertia ($J$) up to ±50% of the nominal value of 0.2 in Table 1.

$$A_1 = \begin{bmatrix} -4 & -0.03 \\ 0.5 & -6.667 \end{bmatrix}, B_1 = \begin{bmatrix} 2 \\ 0 \end{bmatrix};$$
$$A_2 = \begin{bmatrix} -4 & -0.03 \\ 0.75 & -10 \end{bmatrix}, B_2 = \begin{bmatrix} 2 \\ 0 \end{bmatrix};$$
$$A_3 = \begin{bmatrix} -4 & -0.03 \\ 1.5 & -20 \end{bmatrix}, B_3 = \begin{bmatrix} 2 \\ 0 \end{bmatrix};$$
(19)

If the sampling period $h$ is 0.1 sec for the plant, then we get three different state space matrices for the same plant (18) with

$$F_1 = \begin{bmatrix} 0.6703 & -0.0018 \\ 0.0294 & 0.5134 \end{bmatrix}, G_1 = \begin{bmatrix} 0.1648 \\ 0.0035 \end{bmatrix};$$
$$F_2 = \begin{bmatrix} 0.6703 & -0.0015 \\ 0.0378 & 0.3678 \end{bmatrix}, G_2 = \begin{bmatrix} 0.1648 \\ 0.0048 \end{bmatrix};$$
$$F_3 = \begin{bmatrix} 0.6702 & -0.0010 \\ 0.0502 & 0.1353 \end{bmatrix}, G_3 = \begin{bmatrix} 0.1648 \\ 0.0073 \end{bmatrix}.$$
(20)

It is noted that three different discrete state space models are stable in open loop since their eigen-values are inside unit circle. But arbitrary switching between such stable plants may lead to instability as reported in [14] which motivates the design of stabilizing networked predictive controller to tackle arbitrary due to packet losses as well as sudden load change in the plant, representing sudden jerks.

$$\text{eig}(F_1) = 0.67, 0.5137;$$
$$\text{eig}(F_2) = 0.671, 0.368;$$
$$\text{eig}(F_3) = 0.6701, 0.1354$$
(21)

Let $N_{drop} = 3$, i.e. maximum two consecutive packets are considered to be dropped in the simulation. By applying the design procedure first we have to choose a starting point $(P(0), Q(0), K_q(0))$. By using 'YALMIP' Toolbox [13] in the MATLAB environment, we can find out a feasible set of solution satisfying (10), (13) and (15) with the LMI solution and state feedback gains given by (22).

$$P(0) = 10^8 \begin{bmatrix} 2.7140 & 0 & 0 & 0 & 0 & 0 \\ 0 & 2.7140 & 0 & 0 & 0 & 0 \\ 0 & 0 & 2.7140 & 0 & 0 & 0 \\ 0 & 0 & 0 & 2.7140 & 0 & 0 \\ 0 & 0 & 0 & 0 & 2.7140 & 0 \\ 0 & 0 & 0 & 0 & 0 & 2.7140 \end{bmatrix}$$

$$Q(0) = 10^8 \begin{bmatrix} 2.7140 & 0 & 0 & 0 & 0 & 0 \\ 0 & 2.7140 & 0 & 0 & 0 & 0 \\ 0 & 0 & 2.7140 & 0 & 0 & 0 \\ 0 & 0 & 0 & 2.7140 & 0 & 0 \\ 0 & 0 & 0 & 0 & 2.7140 & 0 \\ 0 & 0 & 0 & 0 & 0 & 2.7140 \end{bmatrix}$$

$$K_1(0) = [-4.0718 \quad -0.0745];$$
$$K_2(0) = [0.0019 \quad 0.0281];$$
$$K_3(0) = [0.0006 \quad 0.0132];$$
(22)

We continue the step 2-5 in design procedure. After 10 iterations we obtain an acceptable solution with the following variables:

$$P(10) = \begin{bmatrix} 1.0821 & 0.0062 & 0.0000 & 0.0013 & 0.0000 & 0.0005 \\ 0.0062 & 1.2480 & -0.0018 & -0.0645 & -0.0005 & -0.0204 \\ 0.0000 & -0.0018 & 1.0000 & -0.0007 & 0.0000 & 0.0014 \\ 0.0013 & -0.0645 & -0.0007 & 0.9499 & -0.0009 & -0.0401 \\ 0.0000 & -0.0005 & 0.0000 & -0.0009 & 0.9243 & -0.0015 \\ 0.0005 & -0.0204 & 0.0014 & -0.0401 & -0.0015 & 0.8493 \end{bmatrix}$$

$$Q(10) = \begin{bmatrix} 0.9242 & -0.0047 & 0.0000 & -0.0016 & -0.0000 & -0.0008 \\ -0.0047 & 0.8046 & 0.0014 & 0.0556 & 0.0005 & 0.0219 \\ -0.0000 & 0.0014 & 1.0001 & 0.0008 & -0.0000 & -0.0016 \\ -0.0016 & 0.0556 & 0.0008 & 1.0587 & 0.0011 & 0.0514 \\ -0.0000 & 0.0005 & -0.0005 & 0.0011 & 1.0819 & 0.0019 \\ -0.0008 & 0.0219 & -0.0016 & 0.0514 & 0.0019 & 1.1805 \end{bmatrix}$$

such that these values almost satisfies the condition (14) and the networked robust predictive controller [7], which stabilizes the above system, can be designed easily from the following solution of $K_q$,

$$K_1(10) = [-4.0726 \quad -0.0823];$$
$$K_2(10) = [0.0024 \quad 0.0375];$$
$$K_3(10) = [0.0006 \quad 0.0114];$$
(23)

With initial states $x_0 = [-3 \quad 2]^T$, the simulation results of the DC motor plant for three different value of $J$ are shown under the designed controller in Fig. 3. The simulation studies show that the state variables get stabilized in spite of the arbitrary switching in the plant and network. Three set of state variables in Fig. 3 correspond to the different state matrices as in (20). If the switched plant (19) runs at sampling period 0.2 sec, after 10 iterations we obtain following gains, given by (24). The corresponding time evolution of the stabilized states for the switched load condition is shown in Fig. 4.

$$K_1(10) = [-1.6351 \quad -0.0225];$$
$$K_2(10) = [0.0013 \quad 0.0097]; \quad (24)$$
$$K_3(10) = [0.0001 \quad 0.0015];$$

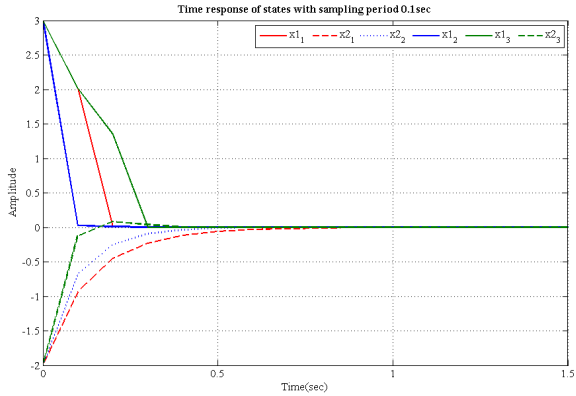

Figure 3. Time response characteristics of the plant for three different value of $J$, operating at a sampling time of 0.1 sec.

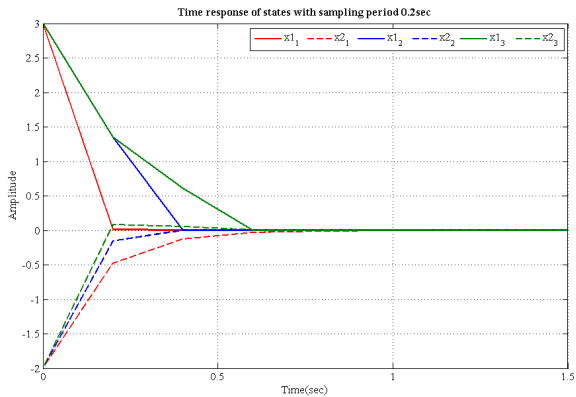

Figure 4. Time response characteristics of the plant for three different value of $J$, operating at a sampling time of 0.2 sec.

As the packet loss process is stochastic in nature, it is not possible to predict at which time instant the packet losses occur. Hence time response can differ from the responses which are shown in Fig. 3 and Fig. 4. But it can be observed that the plants are stabilized within few seconds and the stability is guaranteed in the sense of Lyapunov in spite of having random switching due the network packet losses and random load variation.

## V. CONCLUSION

A predictive networked state feedback controller has been designed to handle arbitrary switching in the plant parameter and packet losses. Lyapunov stability condition yields few set of non-strict LMIs which are solved using the CCL based sequential optimization technique subjected to LMI constraints. Simulations with a DC motor plant have been shown to validate the proposition with arbitrary switching in its load (moment of inertia) along with the switching due to packet losses in the communication network. The simulations are carried out for two different sampling times and the stabilizing controllers are obtained in each case to show wide applicability of the proposed design method. Future scope of research can be directed on extending the proposed concept for event driven controllers as well, with random network induced delays.